\documentclass[11pt,a4paper]{article}
\usepackage[sort]{cite}
\usepackage{url,colortbl}
\usepackage{xcolor}
\usepackage{bm,amsmath,amssymb}
\usepackage[]{graphicx}
\usepackage[margin=2.5cm]{geometry}
\usepackage[ruled,linesnumbered]{algorithm2e}
\usepackage{todonotes}
\graphicspath{{./Images/}}
\usepackage{authblk}
\usepackage{xr}

\newcommand{\ddt}{\frac{\text{d}}{\text{d}t}}
\newcommand{\dd}{\text{d}}

\begin{document}

\title{Optimal control of collective electrotaxis in epithelial monolayers}
\author[1]{Simon F. Martina-Perez\thanks{Corresponding author: martinaperez@maths.ox.ac.uk}}
\author[2]{Isaac B. Breinyn}
\author[3]{Daniel J. Cohen}
\author[1]{Ruth E. Baker}
\affil[1]{Mathematical Institute, University of Oxford, Oxford, United Kingdom}
\affil[2]{Department of Quantitative and Computational Biology, Princeton University, Princeton, NJ, USA}
\affil[3]{Department of Mechanical and Aerospace Engineering, Princeton University, Princeton, NJ, USA}
\date{}
\maketitle
\begin{abstract}
    Epithelial monolayers are some of the best-studied models for collective cell migration due to their abundance in multicellular systems and their tractability. Experimentally, the collective migration of epithelial monolayers can be robustly steered \textit{e.g.} using electric fields, via a process termed electrotaxis. Theoretically, however, the question of how to design an electric field to achieve a desired spatiotemporal movement pattern is underexplored. In this work, we construct and calibrate an ordinary differential equation model to predict the average velocity of the centre of mass of a cellular monolayer in response to stimulation with an electric field. We use this model, in conjunction with optimal control theory, to derive physically realistic optimal electric field designs to achieve a variety of aims, including maximising the total distance travelled by the monolayer, maximising the monolayer velocity, and keeping the monolayer velocity constant during stimulation. Together, this work is the first to present a unified framework for optimal control of collective monolayer electrotaxis and provides a blueprint to optimally steer collective migration using other external cues. 
\end{abstract}

\section{Introduction}\label{chapter3_section:Intro}
Electrotaxis is the process by which eukaryotic cells establish a cell polarity and move directionally in the presence of an electric field \cite{Cohen2014GalvanotacticMonolayers, Allen13, Zajdel2020SCHEEPDOG:Migration}. Electrotaxis is strongest in cellular collectives  due to E-cadherin mediated adhesions  \cite{Wolf2021Short-termDynamics, li13collective}, and displays striking similarities between different cell types and electric field stimulation patterns. Previous studies have focused on using electrotaxis as a means to achieve particular patterns in collective movement \cite{Cohen2014GalvanotacticMonolayers, Zajdel2020SCHEEPDOG:Migration}. For instance, electric fields can be programmed to give bespoke migration patterns \cite{Zajdel2020SCHEEPDOG:Migration},  which are along the field lines in converging and diverging electric fields \cite{Cohen2014GalvanotacticMonolayers}.  A common theme in all of these experiments is that the average direction of collective migration can be accurately controlled during electric field stimulation. The ease with which electric fields can establish directed migration in cellular collectives has led electrotaxis to be accepted as a robust method of steering collective migration \cite{leal23coculture}. The theoretical interest in modelling collective electrotaxis then lies in the ability to answer the following question: Given a desired spatio-temporal movement pattern, how should the electric field be designed to achieve this?

The adoption of electrotaxis as a predictable driver of collective migration has been stalled by a poor understanding of the temporal dynamics of the collective speed during electric field stimulation. For example, Wolf \textit{et al.}~\cite{Wolf2021Short-termDynamics} found that MDCK collectives display a strong, but transient response to the electric field: while the average speed in the direction of the field increases quickly after the electric field is turned on, the collective starts to slow down after approximately one hour, even though the electric field strength remains constant. In general, the current lack of understanding of how the speed of the cellular collective will evolve during stimulation with an electric field poses difficulties in designing bespoke stimulation strategies that can be used to achieve a predictable migration outcome. This difficulty places the question of predicting collective speed during electrotaxis among the most pressing issues in collective electrotaxis. The problem we seek to address in this work is therefore how the collective speed of MDCK monolayers during stimulation with a given electric field can be predicted, and how this knowledge could be exploited to design a stimulation strategy to produce a desired pattern of collective migration?

Both theoretically and experimentally, the question of how to control the temporal dynamics of the collective velocity has scarcely received attention in the literature, even though the fact that the temporal dynamics of electrotaxis are non-trivial is well-documented. Controlling the temporal dynamics of collective velocity remains challenging, since the subcellular processes governing the evolution of cell polarity and active force production in response to stimulation with an electric field remain poorly understood \cite{Allen13, Cohen2014GalvanotacticMonolayers, Wolf2021Short-termDynamics, Zajdel2020SCHEEPDOG:Migration}. At the subcellular level, electrophoresis of charged membrane components creates a cell polarity that activates intracellular signalling pathways that are largely shared with those found in chemotaxis \cite{Allen13, song21pten, Gao15}. Subsequent modelling efforts \cite{nwogbaga2023physical} have been able to explain the directional movement of cells during electrotaxis, but the existing literature is not well positioned to explain the temporal dynamics of cell responses.

In this work, we are interested in constructing a deterministic, continuous-time model to predict the average velocity of the centre of mass of a cellular monolayer in response to stimulation with an electric field. We will consider only uniaxial electric fields, \textit{i.e.,} the field can be described solely by its field strength in the positive or negative direction. Under these assumptions, the goal of this work will be to derive a system of ODEs which predict the collective velocity and the active forces in the monolayer. A significant benefit of using ODE models is that there exists a completely developed theory on how to optimally control ODE solutions according to a given optimisation problem \cite{lenhart07optimalcontrol}. In the context of collective electrotaxis, this means that a well-calibrated and reliable ODE model can be used to solve an optimisation problem, whereby the optimality condition is given by the desired  temporal  pattern of collective movement and the control is the electric field input.

The current scholarship on collective migration of epithelia provides a convenient starting point for the construction of such an ODE model -- see \cite{alert20physical} for a review of continuum models of collective epithelial migration. For example, since the internal signaling pathways for collective electrotaxis are largely shared with chemotaxis, one can begin to model the internal cellular response to an electric field with a well-established continuum model for the response to a chemotactic signal. Erban and Othmer proposed an \textit{adaptation-excitation model} for the chemotactic response of single cells, whereby cells are \textit{excited} when initially exposed to an external signal, and slowly adapt over time \cite{erban2005signal}. In their work, signal transduction is interpreted as ``having two input pathways, an excitatory one that stimulates the [response], and one [...] which, in turn, shuts off the response''.  Similar models are frequently proposed to describe chemotaxis. The advantage of such a model is that it is given by a straightforward system of scalar ODEs that describe the \textit{effective signal} transduced within a cell upon contact with a chemoattractant \cite{erban2005signal}. In this work, we will assume that similar internal signaling dynamics can be used to model the cellular response to an applied electric field. By letting the strength of the active forces in the monolayer be described using the model of Erban and Othmer \cite{erban2005signal}, we can derive a closed system of ODEs to describe the monolayer speed. Using experimental data in conjunction with Bayesian inference, we can interrogate the validity and applicability of using such a system of ODEs to model collective electrotaxis in response to stimulation with an electric field. Put together, the model can be used to develop a framework for designing electric fields that lead to desired outcomes in the form of an optimal control experiment. 

This work will be structured as follows. In Section~\ref{chapter3_section:Data} we present the experimental data of collective electrotaxis that will be used for calibration and validation of the ODE models. Then, in Section~\ref{chapter3_section:AdaptationExcitation}, we derive a simple ODE model that describes the evolution of the tissue velocity. We then use this system of ODEs to predict the optimal control policy when a number of targets are considered, and compare our results to the current standards for electric field stimulation used in the literature. Together, we offer a mathematical modelling framework to describe, understand, and optimise key cellular processes involved in the regulation of collective electrotaxis.

\section{Methods and data} 
\label{chapter3_section:Data}
\begin{figure}
    \centering
    \includegraphics[width=0.75\linewidth]{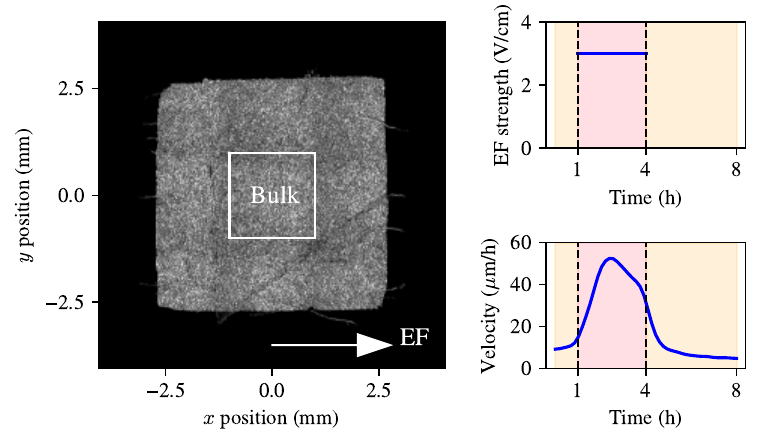}
    \caption{Experimental data of MDCK monolayer electrotaxis with its electric field stimulation protocol. Left panel: phase field image of an MDCK epithelial monolayer indicating direction of the uniaxial electric field and bulk region where the mean velocity is measured. Right panel, top: electric field density during the experiment. Right panel, bottom: bulk velocity corresponding to the electric field trace in the top row.}
    \label{chapter3_fig:data}
\end{figure}

In this section, we present the publicly available experimental data from Wolf~\textit{et al.} \cite{Wolf2021Short-termDynamics}. In brief, the experiments of Wolf \textit{et al.} \cite{Wolf2021Short-termDynamics} involved electrotaxis of 5$\times$5 mm$^2$ MDCK-II epithelial monolayers. In these experiments, a total of nine MDCK epithelial monolayers were grown to confluence and stimulated with an electric field of 3V/cm -- see stimulation trace in Figure~\ref{chapter3_fig:data}. In each experimental replicate, 5$\times$5mm$^2$ square tissues were seeded onto 10cm tissue-culture plastic dishes (CELLTREAT) using stencils of 250$\mu$m-thick silicone elastomer (Bisco HT6240) cut with a hobbyist razor-writer (Silhouette Cricut). Suspensions of MDCK cells were seeded into the stencil patterns at a volume of 10$\mu$L and density of 2.5 ($\pm$0.15) x 10$^6$ cells/mL. Cells were given one hour to adhere before the dish was flooded with media and incubated for 18 hours prior to stencil pull and  insertion into the electric bioreactor. Tissue culture stencils were pulled approximately 1 hour before imaging began. Immediately post-stencil pull, dishes were refilled with fresh culture media followed by direct integration with the bioreactor. The assembled bioreactor is then placed into the microscope, situated within a custom-built cage incubator maintained at 37$^{\circ}$C and perfused with fresh media continuously bubbled with 5\% CO$_2$ using a peristaltic pump (Instech Laboratories) at a rate of 2.5 mL/h. Electrotaxis was induced by custom electro-bioreactors developed by the Cohen group at Princeton University, which are based on the SCHEEPDOG platform~\cite{Zajdel2020SCHEEPDOG:Migration}. These bioreactors provide a continuous flow of media and can be programmed to generate electrical fields with custom electric field densities. 

Velocity data from each of the experiments were acquired using particle image velocimetry (PIV) on phase contrast images,  where the cell nuclei are used as the fiduciary markers.  Local tissue velocities were computed every 10 minutes. The PIV data provide a velocity field at a spatial resolution of 32$\times$32 $\mu$m$^2$, which corresponds roughly to 3$\times$3 cell lengths. For a sample velocity heat map, see Figure~\ref{chapter3_fig:data}. For further details on the PIV data processing and implementation details in FIJI \cite{fiji12} and MATLAB \cite{matlab2022}, we refer the reader to \cite{Wolf2021Short-termDynamics}. Experimental PIV data provides detailed spatiotemporal information on the velocities at a fine spatial resolution. Wolf \textit{et al.} \cite{Wolf2021Short-termDynamics} previously used this detailed spatial data to characterise the responses in different tissue locations during collective electrotaxis. They found that the edges of the MDCK monolayer have a preferred direction of movement, which is outward of the tissue and is only observed within approximately 200$\mu$m of the tissue boundary. This effect is a hallmark of the collective migration of MDCK cells \cite{Wolf2021Short-termDynamics, Tilli18, Cohen2014GalvanotacticMonolayers, alert20physical}. Although some attention has been given to understanding how the preferred migratory direction of the edges interferes with electrotaxis in the literature \cite{Wolf2021Short-termDynamics}, it is not necessary to incorporate detailed mechanisms of edge migration when understanding the average velocity of the centre of mass of the epithelial monolayer. Since the majority of the tissue does not exhibit the preferred direction of migration that the edges show, we restrict our analysis to the \textit{tissue bulk}, the 3$\times$3 mm$^2$ region in the centre of the tissue characterised by Wolf \textit{et al.} \cite{Wolf2021Short-termDynamics} -- for a schematic, see Figure~\ref{chapter3_fig:data}. We take a spatial average of the velocity in the bulk, which yields one-dimensional data traces that represent the monolayer velocity in each of the experiments.

\section{An adaptation-excitation model of collective electrotaxis}
\label{chapter3_section:AdaptationExcitation}
We use a known model for cellular excitation and adaptation proposed by Erban and Othmer \cite{erban2005signal} to model the internal cellular response to electric field. Since all the experiments we study consider a uniaxial electric field, we describe the migration of the monolayer bulk using the $x$ component of the velocity, $v$, in the positive direction of the electric field, which is assumed aligned with the $x$-axis, as shown in Figure~\ref{chapter3_fig:data}. The goal of this section is to connect a model of cellular adaptation and excitation to the temporal evolution of the bulk velocity in response to the electric field and furthermore, to seek to use this model to address the question of how to optimally control the collective migration of MDCK epithelia by varying the field strength.

\subsection{Model derivation}
 We assume that cells experience a viscous friction force per unit mass which is proportional to their velocity with a friction coefficient, $\gamma$, such that the friction force per unit mass is given by $-\gamma v.$ The assumption of linear friction is well-founded for epithelia, since this corresponds to the polymerisation of adhesions on the substrate \cite{alert20physical}.  We assume throughout that there will be a uniaxial electric field in the positive $v$-direction whose magnitude is given by a scalar signal, $s$. In what follows, we will introduce in detail the Erban and Othmer model \cite{erban2005signal} to describe how the electric field creates an intracellular \textit{effective signal} to the cytoskeletal machinery to produce an active force in the direction of the electric field.

The canonical model for excitation-adaptation proposed by Erban and Othmer considers an external stimulus, $s$, an \textit{effective signal}, $s_{\text{eff}}$, and an \textit{inhibitor}, $I$. The effective signal, $s_{\text{eff}}$, is a scalar quantity that describes the extent to which an external signal is transduced within the cell. There is therefore a difference between the constant electric field supplied and the resulting internal signal. The inhibitor, in contrast, represents the strength of the internal signalling pathways that try to inhibit this response to the external stimulus. The dynamics are given by
\begin{align}
    \dot{s}_{\text{eff}} &= \frac{s - (s_{\text{eff}} + I)}{\tau_e}, \label{chapter3_eq:EffectiveSignal}\\
    \dot{I} &= \frac{s - I}{\tau_a}. \label{chapter3_eq:inhibitor}
\end{align}
In this model, one assumes that the excitation timescale, $\tau_e > 0$, is far shorter than the adaptation timescale, $\tau_a > 0$. The different timescales in this system create a characteristic response to a constant signal \cite{erban2005signal}: when the external signal, $s(t)$, is a Heaviside function, the solution of Equations~\eqref{chapter3_eq:EffectiveSignal} and \eqref{chapter3_eq:inhibitor} is of the form
\begin{equation}
    s_{\text{eff}} = \frac{\tau_a}{\tau_a - \tau_e}\left(e^{-t/\tau_a} - e^{-t/\tau_e}\right), \quad I = 1-e^{-t/\tau_a}. \label{chapter3_eq:asymptoticSeff}
\end{equation}
With other words, as $t \to \infty$, the effective signal wanes, \textit{i.e.}, $s_{\text{eff}} \to 0$. In particular, since $\tau_e \ll \tau_a$, 
\begin{equation}
    s_{\text{eff}} \sim C e^{-\tau_a t},
\end{equation}
whenever $t \gg \tau_e$. This implies that on time scales larger than the excitation time, $\tau_e$, the field response decays exponentially, so that a constant electric field will not produce an internal signal that leads to sustained migration at timescales much greater than the initial excitation time, $\tau_e$. In this work, we consider $s$, $s_{\text{eff}}$, and $I$ to be unitless, as we rescale the field strength, $s$, by dividing by the canonical field strength of 3V/cm in the experimental set-up. Finally, we make the assumption that whenever $s \geq 0$, the effective signal, $s_{\text{eff}}$, equally satisfies $s_{\text{eff}} \geq 0$. The reason for making this additional assumption is that Equations~\eqref{chapter3_eq:EffectiveSignal} and~\eqref{chapter3_eq:inhibitor} allow for $s_{\text{eff}}$ to become negative when the field is turned off. MDCK epithelial monolayers are not observed to move in the direction opposite to the electric field when the electric field is turned off, and is not physically realistic. For this reason, going forward, unless explicitly noted, we consider the quantity
\begin{equation}
    s_{\text{eff}}^+ = \max(s_{\text{eff}}, 0).
\end{equation}

To close the system of equations, we now relate the system of effective signal and inhibitor to the velocity. We assume that the active force  per unit mass  in the direction of the electric field is proportional to the effective signal. This assumption follows canonical models in the literature for coupling polarity and force magnitude \cite{alert20physical}, and results in the following expression for the active force per unit mass, $\tilde{F}_{\text{active}}$,
\begin{equation}
    \tilde{F}_{\text{active}} = \alpha s_{\text{eff}}^+,
\end{equation}
where $\alpha > 0$ is a parameter that controls the responsiveness of the bulk to the input signal. Note that, since $s$ is unitless, $\alpha$ is given in units of $\mu$m/h$^2$.  In this section, $\alpha$ is assumed to be a constant, but we will explore different functional forms for the field response in the next section. By then assuming that the resultant force on the bulk is given by the sum of the active force per unit mass and the friction force  per unit mass, one obtains a single equation for the velocity,
\begin{equation}
    \label{chapter3_eq:constantFieldResponse}
    \dot{v} = -\gamma v + \alpha s_{\text{eff}}^+.
\end{equation}

\subsection{Linear stability analysis}
We begin by noting that Equations~\eqref{chapter3_eq:EffectiveSignal}, \eqref{chapter3_eq:inhibitor}, and \eqref{chapter3_eq:constantFieldResponse} can be written as a linear system of equations such that
\begin{equation}
    \label{chapter3_eq:linearSystem}
    \ddt \begin{pmatrix} v \\ s_{\text{eff}} \\ I\end{pmatrix} = 
    \begin{pmatrix}
        -\gamma & \alpha & 0 \\
        0 & -\tau_e^{-1} & - \tau_e^{-1} \\
        0 & 0 & -\tau_a^{-1}
    \end{pmatrix} \begin{pmatrix} v \\ s_{\text{eff}} \\ I\end{pmatrix} + 
    \begin{pmatrix} 0 \\ \tau_e^{-1}s(t) \\ \tau_a^{-1} s(t)\end{pmatrix}.
\end{equation}
We assume that the monolayer is stationary at the beginning of stimulation and the electric field is turned off, so that the initial condition is given by 
\begin{equation}
    v(0) = s_{\text{eff}}(0) = I(0) = 0. \label{chapter3_eq:SimpleInitialCondition}
\end{equation}
Note that the eigenvalues of the matrix in the homogeneous part of the right-hand side of Equation~\eqref{chapter3_eq:linearSystem} are $-\gamma, -\tau_e^{-1}, -\tau_a^{-1}$, making the matrix invertible. At steady state, one has that the left-hand side of Equation~\eqref{chapter3_eq:linearSystem} is zero, and that the steady state solution, $(v^\star, s_{\text{eff}}^\star, I^\star)$, for a constant field strength, $s(t) \equiv s$, is given by taking the inverse,
\begin{equation}
    \begin{pmatrix}
        v^\star\\
        s_{\text{eff}}^\star\\
        I^\star
    \end{pmatrix} =
    \begin{pmatrix}
        -\gamma^{-1} & -\frac{\alpha \tau_e}{\gamma} & -\frac{\alpha\tau_a}{\gamma} \\
        0 & -\tau_e & \tau_a\\
        0 & 0 & -\tau_a
    \end{pmatrix} 
    \begin{pmatrix}
    0 \\ \tau_e^{-1} s \\ \tau_a^{-1} s
    \end{pmatrix} = 
    \begin{pmatrix}
    0 \\
    0 \\
    s
    \end{pmatrix}.
\end{equation}
Therefore, Equation~\eqref{chapter3_eq:linearSystem} has a single steady state. Since this steady state is given by zero velocity and zero effective signal, it can be interpreted as corresponding to a tissue that has fully adapted to the electric field and exhibits no response. Moreover, this steady state is stable since all eigenvalues of the matrix on the right-hand side of Equation~\eqref{chapter3_eq:linearSystem} are strictly negative. This explains the waning response with constant stimulation, as there is no steady state where a non-zero velocity is attained.

\subsection{Bayesian inference of model parameters} \label{chapter3_section:simpleBayesianInference}
Our objective is to calibrate the simple adaptation-excitation model in Equation~\eqref{chapter3_eq:linearSystem} to the experimental bulk velocity data of Wolf \textit{et al.} \cite{Wolf2021Short-termDynamics} by performing Bayesian inference on the unknown model parameters $\gamma, \alpha, \tau_e, \tau_a$ from the available bulk data. 

\begin{figure}
    \centering
    \includegraphics[width=0.33\linewidth]{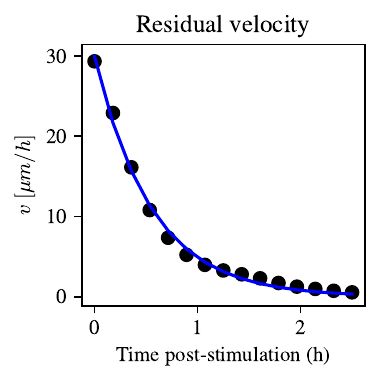}
    \caption[Experimental data of bulk velocity decay post-stimulation together with least-squares fit of exponential decay.]{Experimental data of bulk velocity decay post-stimulation with an electric field pulse of 3 V/cm (black dots) together with least-squares fit of exponential decay in the form $C e^{-\gamma t}$ (solid blue line). The least-squares solution gives an excellent fit to the data and can be used to identify the decay rate, $\gamma$.}
    \label{chapter3_fig:VelocityDecay}
\end{figure}

First, we note that the velocity decay parameter, $\gamma$, can be estimated directly from the data, since Equation~\eqref{chapter3_eq:constantFieldResponse} predicts that, given an initial velocity, $v_0$, the bulk velocity will decay exponentially in the absence of an electric field. Since the experimental data contain information on the bulk velocity after  stimulation with the electric field is turned off,  it follows that the observed velocity decay can be used to immediately infer the decay rate, $\gamma$. From these considerations,  we seek to fit an exponential of the form 
\begin{equation}
    v(t) = C e^{-\gamma (t-t_{\text{end}})},
\end{equation}
to the bulk velocity data in the 2.5 hours after stimulation with the electric field is turned off. Here, $t_{\text{end}}$ is the time at which the electric field is switched off, and $C$ is given by 
\begin{equation}
    C = v(t_{\text{end}}).
\end{equation}
We perform a least-squares fit for the decay rate, $\gamma$, to give an estimate of $\gamma = 1.765$h$^{-1}$. Figure~\ref{chapter3_fig:VelocityDecay} shows that the fitted exponential decay curve is in excellent agreement with the experimental data after pulse stimulation. This motivates us to take the value of $\gamma$ as fixed going forward. The benefit of directly estimating $\gamma$ from data is that it reduces the dimensionality of the parameter space, thus simplifying the inference of the remaining parameters.

\begin{figure}
    \centering
    \includegraphics[width=0.75\linewidth]{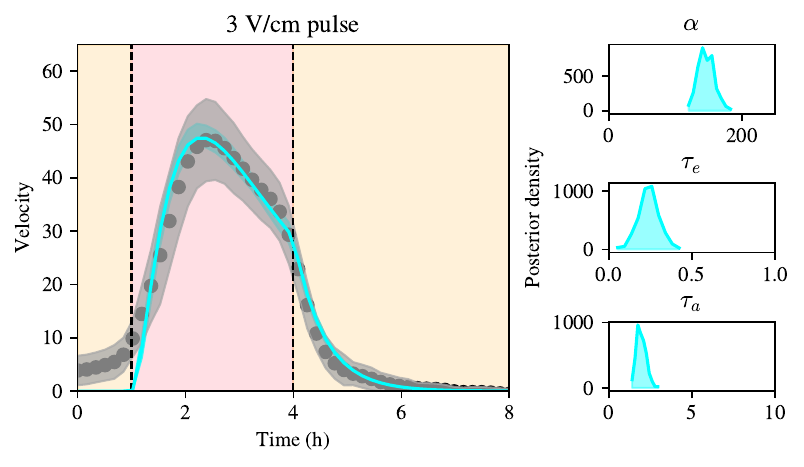}
    \caption{Bayesian inference on a simple model of adaptation and excitation given by Equation~\eqref{chapter3_eq:linearSystem} using experimental data from an electrotaxis experiment. Left: bulk velocity using a constant field of 3V/cm. Posterior mean of the model  together with a confidence interval between the 5\% and 95\% quantiles of the model posterior predictions  in cyan, experimental data and their confidence interval in grey.  Notice the small width of the posterior predictive intervals. Right panels, from top to bottom: posterior distributions for model parameters, $\alpha$, $\tau_e, \tau_a$, respectively.}
    \label{chapter3_fig:BayesianInferenceSimple}
\end{figure}

We now keep $\gamma = 1.765$h$^{-1}$ fixed and perform Bayesian inference of the remaining parameters, $\alpha$, $\tau_e$ and $\tau_a$. We use Markov chain Monte Carlo with a Haario-Bardenet adaptive covariance and four chains under the assumption of a Gaussian error implemented in Python with the PINTS pacakge \cite{clerx19pints}. As a metric of convergence we use the $\hat{R}$ statistic \cite{clerx19pints}, which summarises mixing and stationarity of the chains. We use the reference value of $\hat{R} = 1.05$ for four chains \cite{vehtarh21MCMC} and set a maximum number of MCMC iterations to $2\cdot 10^4$. We use as data the velocity profiles from the 3V/cm pulse experiments from Wolf \textit{et al.} \cite{Wolf2021Short-termDynamics}. The MCMC procedure yields a well-identified posterior distribution, with marginal posterior means 149.92 $\mu$m/h$^2$,  0.260h, and 2.038h, for $\alpha$, $\tau_e$ and $\tau_a$, respectively. These posterior means are in good agreement with previous intuition that the time scale for adaptation is longer than the time scale for excitation, as well as with the data, suggesting that the bulk slows down on a timescale of the same order of magnitude as $\tau_a$. Plots of the marginal posterior distributions of the model parameters are shown in Figure~\ref{chapter3_fig:BayesianInferenceSimple}. These show that the parameters can be confidently identified from the data, given their well-defined posterior distributions. 

To assess the predictive capability of the model given the posterior distribution of the model parameters, we compute the posterior predictive intervals of the model, \textit{i.e.} a 95\% confidence interval for the model outputs given the posterior distribution of the model parameters. These posterior predictive intervals are shown in Figure~\ref{chapter3_fig:BayesianInferenceSimple}. It can be seen that the posterior predictive intervals are tightly concentrated around the mean, showing little model uncertainty in predicting the experimental data, given the posterior distribution of the model parameters. By overlaying the posterior predictive intervals for the model onto the experimental data, we observe that the model predictions are in good agreement with the experimental data and lie comfortably within the standard deviation of the experimental data. We conclude that the adaptation-excitation model presented in this section, when calibrated to the experimental data, is able to recapitulate the phenomenon of attenuating velocity during electrotaxis to the field when the electric field is held constant. 
\subsection{Optimal control} \label{chapter3_section:optimalControlDistanceSimple}
Having established that excitation-adaptation dynamics can be used to describe the velocity of the tissue bulk during electrotaxis, we turn to the question of designing a stimulation protocol that is in some way optimal for the experiment. We previously established that the question of how to design electric field stimulation protocols has received little attention due to the difficulties in predicting tissue response \textit{a priori}. Instead, experiments have been carefully calibrated through trial and error by the experimental laboratories that have performed them over the past few years. At this point, it is worth noting that these experiments are costly and time-consuming, making a grid search or other brute-force methods to determine optimal temporal stimulation patterns infeasible.

In this section, we will consider the problem of predicting an optimal pattern of stimulation using the adaptation-excitation model. We will consider that the optimal control problem can be cast as an optimization problem of a functional, $E$, that depends continuously on the velocity as well as the signal,
\begin{equation}
    E(v,s) = \int_0^T f(v,s) \dd t,
\end{equation}
and that there is a fixed start and end time of stimulation, given by $0$, $T$, respectively. We need to penalise for the total amount of electric current administered during the experiment, as applying large currents throughout the cell medium for large amounts of time yields damage to the tissue and the experimental set-up. For now, we will incorporate this  by requiring that a proxy of the total amount of charge delivered to the tissue during stimulation remains constant,  \textit{i.e.} 
\begin{equation}
    \label{chapter3_eq:chargeBudget}
    \int_0^T s(t)^2 \dd t= s_{\text{budget}}.
\end{equation}
Later in this section, we will investigate the role of introducing lower and upper bounds on the field strength instead. Note also that in Equation~\eqref{chapter3_eq:chargeBudget} we have introduced a quadratic to account for the fact that the electric field can be positively or negatively charged, and that fields of both signs will affect the tissue. 

Under the condition of Equation~\eqref{chapter3_eq:chargeBudget}, we set up the first optimal control problem to maximise the total distance travelled by the tissue. This is a straightforward question to ask when one is concerned with increasing the controllability of collective migration. We therefore consider the optimization problem
\begin{equation}
    \max_{S} \int_0^T v(t)\dd t.
\end{equation}
This optimisation problem is then subject to the condition in Equation~\eqref{chapter3_eq:chargeBudget}. In optimal control theory, this is known as an \textit{isoperimetric problem}. In this formulation, we introduce a new variable, 
\begin{equation}
    z(t) = \int_0^t s(t)^2 \text{d}t,
\end{equation}
such that the problem can be described as
\begin{align}
    \max_S E &= \int_0^T v \dd t, \quad \text{subject to}\\
    \dot{z}(t) &= s(t)^2, \\
    v(0) &= 0,\\
    z(0) &= 0, \quad z(T) = S_{\text{budget}}.
\end{align}
Recall that the isoperimetric contraint above has been introduced to enforce a physical limit on the amount of charge to which cells in the monolayer are exposed. One might also interpret the isoperimetric constraint as a manner of interrogating how, given a fixed amount of charge, one might optimally distribute the delivery of this charge so that the total distance travelled by the monolayer is optimal. To make direct comparisons with the experiments performed in Wolf \textit{et al.} \cite{Wolf2021Short-termDynamics}, we wish to make the stimulation pattern have the same integrated charge for constant stimulation at 3 V/cm, so we set $s_{\text{budget}} = 27$V$^2$h/cm$^2$. For this problem the Hamiltonian is given by
\begin{equation}
\begin{split}
    H &= v + \lambda_v(t)\dot{v} + \lambda_{s_{\text{eff}}}(t) \dot{s}_{\text{eff}} + \lambda_I \dot{I} - \mu_z \dot{z}\\
    &= v + \lambda_v(t) \left[ -\gamma v + \alpha s_{\text{eff}}\right] + \tau_e^{-1}\lambda_{s_{\text{eff}}}(t)\left[s - s_{\text{eff}} - I \right] + \tau_a^{-1}\lambda_I(t)\left[s-I\right] - \mu_z S^2,\\
\end{split}
\end{equation}
where $\partial\lambda_\bullet/\partial t = - \partial H/\partial\bullet$, such that 
\begin{align}
     \frac{\dd \lambda_v}{\dd t} &= - \frac{\partial H}{\partial v} = -1 + \gamma \lambda_v, \label{chapter3_eq:adjointV}\\
     \frac{\dd \lambda_{s_{\text{eff}}}}{\dd t} &= - \frac{\partial H}{\partial s_{\text{eff}}} = -\alpha \lambda_v + \tau_e^{-1} \lambda_{s_{\text{eff}}}\label{chapter3_eq:adjointSeff},\\
     \frac{\dd \lambda_{I}}{\dd t} &= - \frac{\partial H}{\partial I} = \tau_e^{-1} \lambda_{s_{\text{eff}}} + \tau_a^{-1} \lambda_I,\label{chapter3_eq:adjointI}\\
     \frac{\dd \mu}{\dd t} &= -\frac{\partial H}{\partial z} = 0 \label{chapter3_eq:adjointMu},
\end{align}
subject to the terminal conditions $\lambda_\bullet(T) = 0$. Additionally, $\mu = \mu_\star$ is a constant. The optimality condition is given by
\begin{equation}
    0 = \frac{\partial H}{\partial S} = \tau_e^{-1} \lambda_{s_{\text{eff}}} + \tau_a^{-1}\lambda_I - 2\mu_\star S. \label{eq:optimalityCondition}
\end{equation}
The solution to Equation~\eqref{eq:optimalityCondition} yields the optimal stimulation pattern, $S^\star$, which is given by
\begin{equation}
    S^\star = \frac{1}{2\mu_\star}(\tau_e^{-1} \lambda_{s_{\text{eff}}} + \tau_a^{-1}\lambda_I) = \frac{\tau_e^{-1}}{2\mu_\star}\left(\lambda_{s_{\text{eff}}} + \frac{\tau_e}{\tau_a}\lambda_I\right) = \frac{\tau_e^{-1}}{2\mu_\star}\left(\lambda_{s_{\text{eff}}} + \epsilon\lambda_I\right), \label{chapter3_eq:OptimalStimApprox}
\end{equation}
where $\epsilon = \tau_e/\tau_a$. To find the optimal stimulation pattern, $S^\star$, one must seek to solve the linear inhomogeneous system of equations for $\lambda_\bullet$. Note that Equation~\eqref{chapter3_eq:adjointV}, the equation for $\lambda_v$,  decouples from the other adjoint equations,  so we can solve directly to find
\begin{equation}
    \lambda_{v}\left(t\right) = \frac{{\left(\gamma \lambda_{v}\left(0\right) - 1\right)} e^{\gamma t}}{\gamma} + \frac{1}{\gamma}.
\end{equation}
Applying the terminal condition, $\lambda_v(T) = 0$, yields that $\lambda_v(0) = -(e^{-\gamma T}-1)/\gamma$, and so we have
\begin{equation}
    \lambda_{v}\left(t\right) = (e^{\gamma(t-T)}-1)/\gamma.
\end{equation}
Direct substitution of the solution for $\lambda_v$ into Equation~\eqref{chapter3_eq:adjointSeff}, the equation for $\lambda_{s_{\text{eff}}}$, then yields
\begin{equation}
    \begin{split}
            \lambda_{s_{\text{eff}}} &= \frac{\alpha}{\gamma(\gamma - \tau_e^{-1})} - \frac{\alpha}{\tau_e^{-1}(\gamma - \tau_e^{-1})} + c_1 e^{\tau_e^{-1}t} - \frac{\alpha e^{\gamma(t-T)}}{\gamma(\gamma - \tau_e^{-1})}\\
    &= -\frac{\alpha}{\tau_e^{-1}\gamma} + c_1 e^{\tau_e^{-1}t} - \frac{\alpha e^{\gamma(t-T)}}{\gamma(\gamma - \tau_e^{-1})}.
    \end{split}
\end{equation}
We can now fit the terminal condition, $\lambda_{s_{\text{eff}}}(T) = 0$, such that
\begin{equation}
    0 = \lambda_{s_{\text{eff}}}(T) = -\frac{\alpha}{\tau_e^{-1}\gamma} + c_1 e^{\tau_e^{-1}T} - \frac{\alpha}{\gamma(\gamma - \tau_e^{-1})},
\end{equation}
which gives
\begin{align}
    c_1 = e^{-\tau_e^{-1}T}\left(\frac{\alpha}{\tau_e^{-1}\gamma} + \frac{\alpha}{\gamma(\gamma - \tau_e^{-1})}\right) = \frac{\alpha\gamma}{\tau_e^{-1} (\gamma - \tau_e^{-1})}e^{-\tau_e^{-1}T}.
\end{align}
Therefore,
\begin{equation}
    \lambda_{s_{\text{eff}}} = \frac{\alpha\gamma}{\tau_e^{-1}(\gamma - \tau_e^{-1})} \left(e^{\tau_e^{-1}(t-T)} - \frac{\tau_e^{-1}}{\gamma^2}e^{\gamma(t-T)} \right)-\frac{\alpha}{\tau_e^{-1}\gamma}.
\end{equation}
We can then find $\lambda_I$ by solving Equation~\eqref{chapter3_eq:adjointI},
\begin{equation}
    \lambda_I = e^{-\tau_a^{-1}t}\int_0^t \lambda_{s_{\text{eff}}}(\xi) e^{\tau_a^{-1}\xi}\dd \xi + C_I e^{\tau_a^{-1}t},
\end{equation}
where $C_I$ is determined by setting the terminal condition, $\lambda_I(T) = 0$, which gives
\begin{equation}
    C_I = -e^{-\tau_a^{-1}T}\int_0^T \lambda_{s_{\text{eff}}}(\xi) e^{\tau_a^{-1}\xi}\dd \xi.
\end{equation}
Therefore,
\begin{equation}
    \label{chapter3_eq:optimal_simple}
    S^\star = \frac{\tau_e^{-1}}{2\mu^\star} \lambda_{s_{\text{eff}}} + \epsilon\frac{\tau_e^{-1}}{2\mu^\star}\left[e^{-\tau_a^{-1}t}\int_0^t \lambda_{s_{\text{eff}}}(\xi) e^{\tau_a^{-1}\xi}\dd \xi + C_I e^{\tau_a^{-1}t} \right] . 
\end{equation}
To determine the solution of the system, it remains to find the constant $\mu^\star$ such that the optimal solution integrates to $S_{\text{budget}}$. We use the marginal posterior means for the model parameters $\alpha, \tau_e, \tau_a$ to first compute $S^\star$ numerically and use numerical integration to determine the integral during the stimulation window. 

We show a prediction for the optimal stimulation pattern in Figure \ref{chapter3_fig:OptimalShortStim}. Interestingly, the optimal stimulation pattern begins at a nonzero electric field value and increases for the first half of the stimulation window, when it exceeds the constant electric field strength of 3V/cm, before dropping below that field strength in the second half of the stimulation window. The optimal stimulation pattern found in Equation~\eqref{chapter3_eq:optimal_simple} shows only a modest increase in the distance travelled by the bulk of the monolayer compared to the experimental data, since the integrated numerical solution of the velocity is 2.17\% greater than the distance travelled using a constant electric field. This reflects the fact that electrotaxis experiments have been iterated many times experimentally. Strikingly, the optimal stimulation pattern also results in a bulk velocity that still exhibits a slowdown during stimulation, which runs counter to the view in the electrotaxis community that the slowdown of the bulk during stimulation represents nonoptimality of the stimulation procedure.

We incorporate parameter and prediction uncertainty by computing the posterior distribution over the optimal control solution. We take $2\cdot10^3$ samples from the MCMC posterior distribution and for each sample compute the optimal stimulation protocol according to Equation~\eqref{chapter3_eq:optimal_simple} using that given parameter combination and recomputing the normalising constant, $\mu^\star$, for each new parameter combination to obtain a normalised density.  We again compute the 5\% and 95\% percentiles of the velocity distribution at each time point and plot them in Figure~\ref{chapter3_fig:OptimalShortStim}. We see that the posterior distribution is tight around the posterior mean, which confirms that the above analysis holds true across the entire posterior distribution.

\subsection{Optimal control for the terminal velocity}
\begin{figure}
    \centering
    \includegraphics[width=\linewidth]{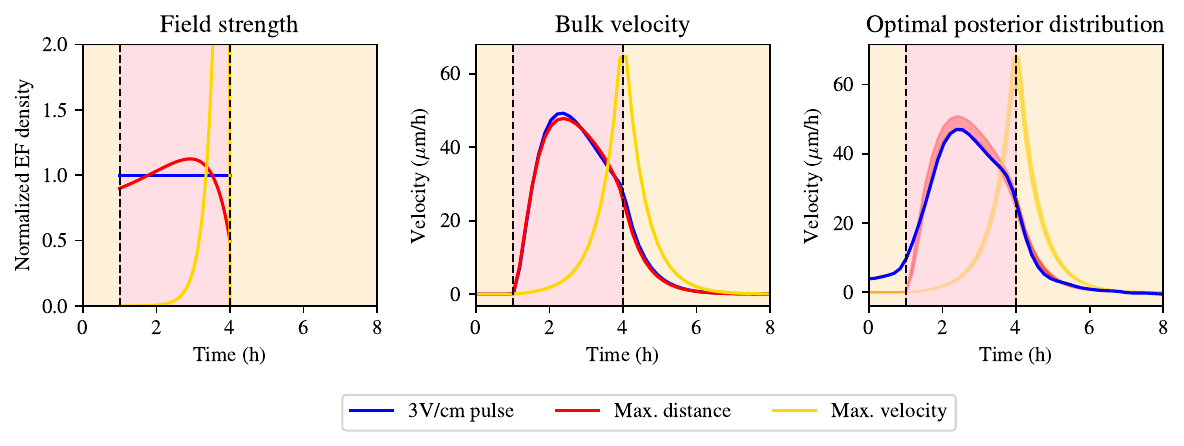}
    \caption{Stimulation patterns for optimal control of collective electrotaxis. In all plots, red corresponds to maximum tissue displacement, pink corresponds to maximum terminal velocity. Green curves represent 3V/cm electric field pulse stimulation pattern used experimentally. Left: the different stimulation patterns in normalized units. Middle: corresponding velocity curves using the different stimulation patterns. Right: data from 3V/cm pulse experiment (dashed line, green), compared to posterior distribution for optimal solution for distrance travelled (red) and terminal velocity (pink).}
    \label{chapter3_fig:OptimalShortStim}
\end{figure}

Optimising for the total distance travelled predicts that there is only a very moderate improvement in the optimised outcome compared to stimulation with a constant electric field of 3V/cm.  A striking feature of the optimal stimulation curve shown in Figure~\ref{chapter3_fig:OptimalShortStim} is that the velocity curve under the optimal stimulation protocol, like the pulse stimulation protocol, displays a diminishing response to the field at later times. Therefore, we pose the additional question of how to maximise the terminal velocity of the tissue, in other words: How can the stimulation be designed so that the bulk velocity at the terminal time, $T$, is maximal? \textit{A priori}, one would expect this to be a different stimulation pattern than the optima found previously. We change the objective function so that we maximise the terminal velocity, $v(T)$,
\begin{equation}
    E(v) = \int_0^T \dot{v} \dd t = \int_0^T [-\gamma v + \alpha s_{\text{eff}}] \dd t.
\end{equation}
Note that this change in objective function only changes the equations for $\dd \lambda_v/\dd t$ and $\dd \lambda_{s_{\text{eff}}}/\dd t$ in the adjoint equations, Equations \eqref{chapter3_eq:adjointV}-\eqref{chapter3_eq:adjointMu}, since the only additional terms introduced into the Hamiltonian include the velocity, $v$, and the effective signal, $s_{\text{eff}}$. We obtain
\begin{align}
    \frac{\dd \lambda_v}{\dd t} &=  \gamma + \gamma \lambda_v, \\
    \frac{\dd \lambda_{s_{\text{eff}}}}{\dd t} &= -\alpha\lambda_v + \tau_e^{-1}\lambda_{s_{\text{eff}}},
\end{align}
such that, directly fitting the terminal condition, $\lambda_v(T) = 0$, we have 
\begin{equation}
    \lambda_v = e^{\alpha(t-T)} -1,
\end{equation}
which we substitute into Equation~\eqref{chapter3_eq:adjointSeff} to find 
\begin{equation}
    \lambda_{s_{\text{eff}}} = \frac{\alpha}{\alpha-\tau_e^{-1}}\left( e^{\tau_e^{-1}(t-T)} - e^{\alpha(t-T)}\right).
\end{equation}
The same computation as in Equation~\eqref{chapter3_eq:optimal_simple} now yields predictions of the optimal electric field stimulation protocol, which results in a terminal velocity of 62.9 $\mu$m/h compared to 52.5$\mu$m/h for a constant field. The electric field that optimises the terminal velocity is shown in Figure~\ref{chapter3_fig:OptimalShortStim}. Note that the field strength is monotonically increasing to its maximum intensity at the end of stimulation. We obtain a higher terminal velocity than in the other experiments, at the cost of a very high field strength at the end of stimulation, approximately 6 V/cm, which is physiologically and experimentally feasible, albeit not for long periods of time. The fact that the optimal control problem for the terminal velocity yields a ramping pattern suggests that total distance travelled and terminal velocity are two incompatible requirements to fulfill simultaneously, given the model and evidence presented.  Finally, to address parameter and prediction uncertainty, we repeat the computation of posterior distributions over the velocities, as done in Section~\ref{chapter3_section:optimalControlDistanceSimple}, which we display in Figure~\ref{chapter3_fig:OptimalShortStim}. Again, the posterior distribution of the velocities is tight around the posterior mean.

\subsection{Optimal control for constant velocity} \label{chapter3_section:simpleOptimalControl}
The optimal control problems for the maximum distance travelled and the maximum terminal velocity seem to suggest that maximising the distance comes at the cost of a decrease in velocity during stimulation, while optimising the terminal velocity does not maximise the total distance travelled. The finding that the optimal stimulation pattern for total distance travelled does not yield a nondecreasing bulk velocity during the course of stimulation with the electric field runs counter to intuition in the electrotaxis community, as the current opinion in the field tends to view a drop in bulk velocity as a source of nonoptimality in stimulation. From an experimental viewpoint, controlling field strength so that the migratory velocity is constant is attractive, since electric fields are often used for `cruise control'. Therefore, we set up a final optimal control problem designed to keep variations in the velocity minimal. We do this by once more modifying the functional so that we seek solutions that minimise
\begin{equation}
    E(\dot{v}) = \int_0^T \dot{v}^2 \dd t = \int_0^T (-\gamma v + \alpha s_{\text{eff}})^2\dd t \label{chapter3_eq:constantVelProblem}.
\end{equation}
As in the previous section, the only adjoint equations affected are those for $\lambda_v$ and $\lambda_{s_{\text{eff}}}$. These become
\begin{align}
    \frac{\dd \lambda_v}{\dd t} &= 2\gamma(- \gamma v + \alpha s_{\text{eff}}) + \gamma \lambda_v, \label{chapter3_eq:constantVelocityAdjointV}\\
    \frac{\dd \lambda_{s_{\text{eff}}}}{\dd t} &= -2\alpha (-\gamma v + \alpha s_{\text{eff}}) - \alpha \lambda_v + \tau_e^{-1} \lambda_{s_{\text{eff}}}. \label{chapter3_eq:constantVelocityAdjointSeff}
\end{align}
In addition, we impose a terminal condition for the velocity, so that we prescribe a desired velocity at the end of stimulation, \textit{i.e.}, $v(T) = v^\star$ for some pre-determined constant, $v^\star$,  which can be set by the practitioner in accordance with a practical need.  Together with Equation~\eqref{chapter3_eq:linearSystem}, Equations \eqref{chapter3_eq:constantVelocityAdjointV}, \eqref{chapter3_eq:constantVelocityAdjointSeff} and the usual transversality conditions, $\lambda_\bullet(T) = 0$, the optimisation of the function in Equation~\eqref{chapter3_eq:constantVelProblem} creates a nonlinear first-order boundary value problem (BVP). The complicated coupling between the equations requires a numerical solution to the problem. Here, we set $v^\star$ so that it is equal to the average bulk velocity during stimulation with a single pulse of 3 V/cm.

For numerical solution of the BVP, we use the bvp4c scheme of Kierzenka \textit{et al.} implemented in Python through the SciPy package~\cite{virtanen20scipy}. This scheme implements a fourth-order collocation algorithm with control of the residuals and uses a damped Newton method with an affine-invariant criterion function. For the implementation, we set a maximum relative tolerance of $10^{-3}$, an absolute boundary value tolerance of $10^{-6}$ and specify a maximum of $10^6$ nodes for the collocation algorithm.
\begin{figure}
    \centering
    \includegraphics[width=\linewidth]{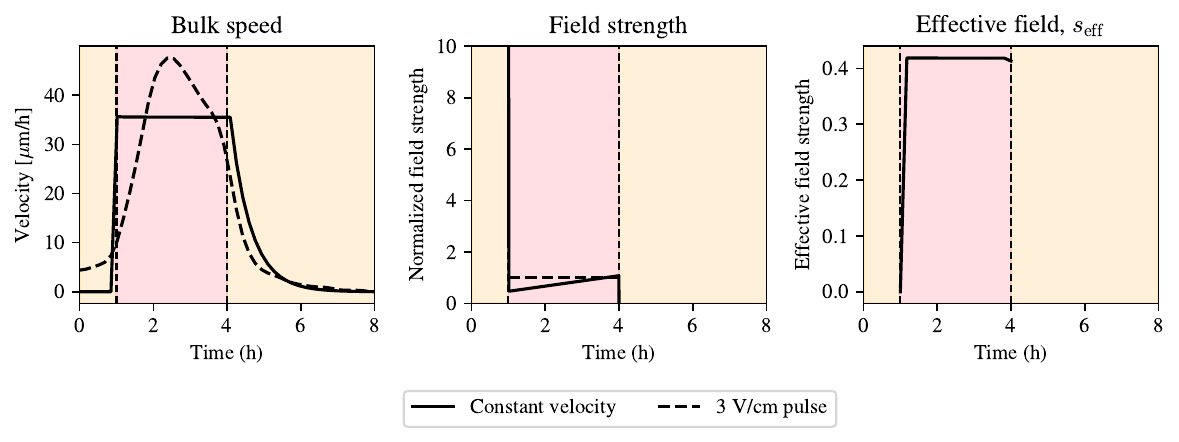}
    \caption{Solution to optimal control problem for constant bulk speed during stimulation with an electric field. Left: bulk speed for constant velocity (solid line) plotted alongside experimental data with constant electric field stimulation protocol (dashed line). Middle: numerical solution of the optimal control problem for constant velocity (solid line) plotted against reference normalized electric field strength (dashed line). Right: effective signal, $s_{\text{eff}}$.}
    \label{chapter3_fig:ConstantVelocityOptimalControl}
\end{figure}

\begin{figure}
    \centering
    \includegraphics[width=\linewidth]{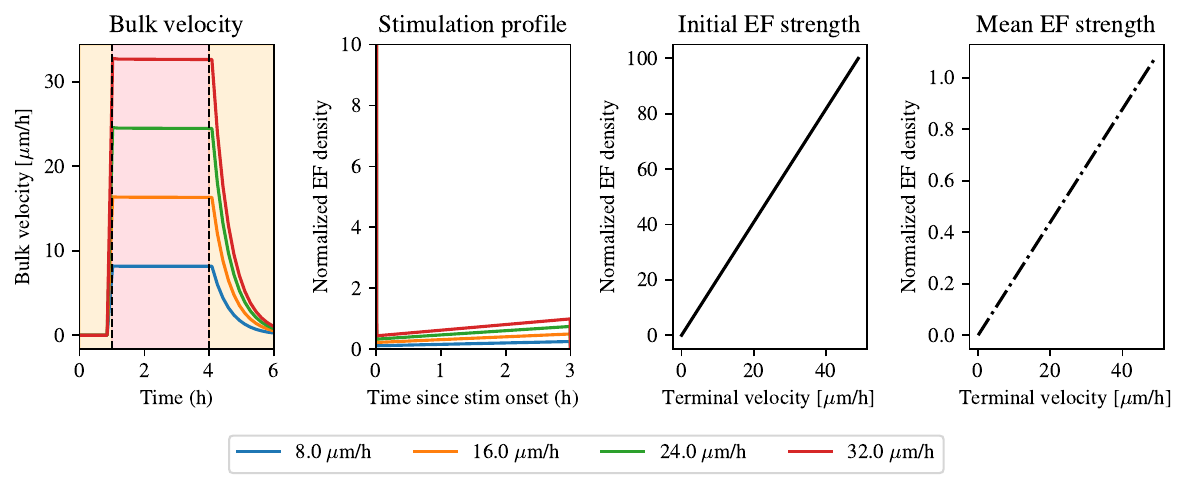}
    \caption{Stimulation profiles and electric field strengths for varying terminal velocities. From left to right. Bulk velocity when using optimal control solution for constant stimulation using different values of terminal velocity (see legend). Optimal stimulation profile corresponding to the different terminal velocities in the left panel. Electric field strength at the start of stimulation. Electric field strength averaged during stimulation. Note the difference in the magnitude of the axes.}
\label{chapter3_fig:VaryingConstantVelocityOptimalControl}.
\end{figure}

Figure~\ref{chapter3_fig:ConstantVelocityOptimalControl} shows that the electric field profile needed to produce a constant velocity profile is very strong initially and rapidly decays to create a constant effective signal, $s_{\text{eff}}$. In fact, the field strength predicted here is 75 V/cm, which is not physiologically realistic, as it would lead to toxicity and rapid cell death. This finding suggests that given the physiological constraints of the problem, it is not possible to design an optimal control experiment that creates a constant speed profile with the same average velocity as that arising from pulse stimulation. This large initial peak in electric field strength arises from the requirement that the tissue hits a constant velocity quasi-instantly upon stimulation. This finding invites the question of which constant velocity values are feasible to achieve given that the solution profile needs to be realistic, \textit{i.e.} given the physical constraints on the electric field strength. To this end, we perform a grid search, whereby we solve the optimal control problem for a range of target velocities, $v^\star$, ranging from $0$ to 40 $\mu$m/h. For each of these terminal velocities, we solve the BVP for the optimal stimulation protocol and compute the resulting bulk velocities and optimal stimulation strategy. We summarise the results in Figure~\ref{chapter3_fig:VaryingConstantVelocityOptimalControl}.

Figure~\ref{chapter3_fig:VaryingConstantVelocityOptimalControl} shows that at each of the values for the terminal velocity, $v^\star$, the model predicts that an optimal stimulation strategy can be designed so that the bulk moves at this constant speed. Each of these stimulation profiles has the same shape, whereby a very large current is applied at the beginning of the experiment, which slowly decreases during the course of the experiment. We investigate the relationship between the desired terminal velocity and the field strength at the onset of stimulation as required by the model. We see that there is a linear increase in the normalised electric field density with $v^{\star}$ and that the initial electric field strength is prohibitively large for velocities higher than 5 $\mu$m/h. This is surprising and experimentally frustrating given that the mean electric field strength predicted by the model only shows a marginal increase with the desired terminal velocity, whereby all terminal velocities experience a normalised electric field strength that is physiologically tolerable. We conclude, thus, that designing a realistic electric field to maintain the tissue bulk at constant velocity, while keeping the electric field strength within realistic bounds, is not possible. 

\begin{figure}
    \centering
    \includegraphics[width=\linewidth]{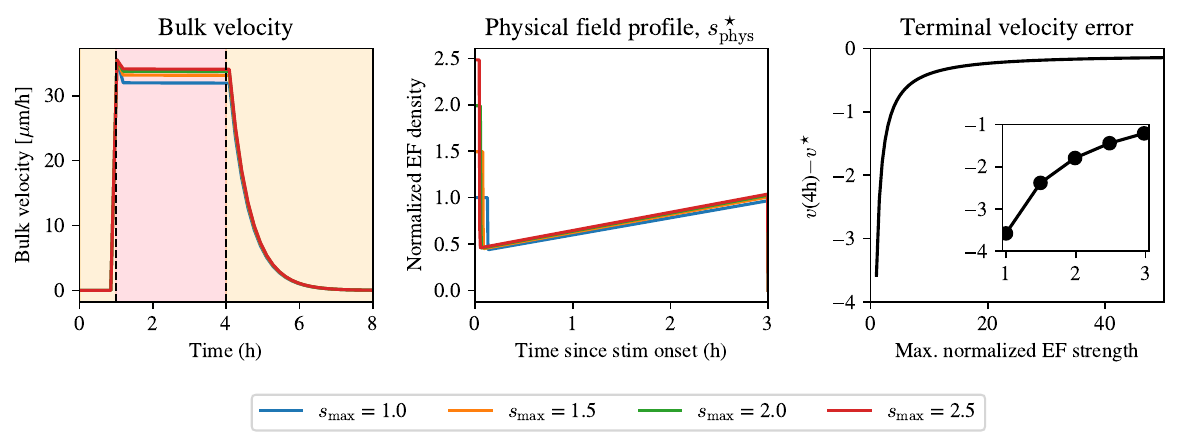}
    \caption[Optimal stimulation to keep the velocity constant during stimulation.]{Optimal stimulation to keep the velocity constant during stimulation. Left: bulk velocity using different maximum electric field strengths (see legend). Middle: physical electric field profile showing similar stimulation protocols. Right: terminal velocity error decreases when the maximum normalized electric field strength increases.}
    \label{chapter3_fig:Clip_Optimal_Control}
\end{figure}

Since the source of the unphysically strong electric field at the start of stimulation is the requirement that the field immediately reaches the desired velocity, we ask if we can stimulate ignoring the non-realistic initial prediction by setting a maximum field strength, since the velocity in the period after the onset of stimulation is indeed constant. We expect this might come at the cost of reaching the target velocity at the beginning of stimulation. We allow for the field to have a field strength of at most 9V/cm, which is around the limit at which electric fields are tolerated \textit{in vitro}. For the optimal control problem in Equation~\eqref{chapter3_eq:constantVelProblem}, we require that the \textit{physical} optimal solution, $s^\star_{\text{phys}}$, satisfies
\begin{equation}
    s_{\text{phys}}^\star = \min\left(s^\star, s_{\max}\right),
\end{equation}
where $s_{\max}$ is a parameter we can vary. By considering $s_{\max}$ across a range between 3 and 9 V/cm, we solve the optimal control problem and show the results in Figure~\ref{chapter3_fig:Clip_Optimal_Control}. We set $v^\star$ as the mean velocity in the 3V/cm pulse experiment and now vary $s_{\max}$. While the solutions are qualitatively similar and the velocity is constant during most the experiment, the solutions are not identical. A closer inspection of the terminal velocity error shows that solutions with a weaker initial field have a larger error at the end, implying that the large initial pulse is what is necessary to create the terminal velocity, but it is the temporal evolution of the field which is necessary to create a plateau. While this error decreases rapidly for small values of the maximum field strength, $s_{\max}$, a close agreement with the desired cruise velocity is only achieved, once again, with unfeasibly high electric field strengths.

\subsection{Constant velocity during a small time interval}
When the goal was to optimise to keep the velocity constant, as in the previous section, we observed that the electric field profile had a non-physically high field density at the start of stimulation, making this optimal control strategy infeasible in practise. This comes from requiring that the velocity hits the target velocity, $v^\star$, quasi-instantly upon the start of stimulation. Here, we ask the question of how solutions change when instead the velocity is only required to be constant on an interval $[t_1, t_2]$ with $0 < t_1 < t_2 < T$. Here, we will explore how the optimal electric field profile is affected by this requirement. We will also study how solutions are affected by altering the values for the start and end of stimulation, $t_1$ and $t_2$, respectively.

The functional in Equation~\eqref{chapter3_eq:constantVelProblem} needs to be modified to address the optimality condition that the velocity is constant on the interval. This is because minimising the square of acceleration $\dot{v}^2$ as in Equation~\eqref{chapter3_eq:constantVelProblem} leads to a trivial solution for the velocity: when the velocity is assumed to be zero by the initial conditions and the velocity is continuous, the optimal solution will be zero throughout the stimulation time frame. Therefore, one must consider a functional that directly relates the velocity during the interval $[t_1, t_2]$ to the target velocity. Such a functional can be defined as
\begin{equation}
    I(v) = \int_0^T \mathbb{I}(t \in [t_1, t_2])\cdot(v-v^\star)^2 \dd t,
\end{equation}
where $\mathbb{I}$ is an indicator function. This indeed restricts the constant velocity requirement to the desired interval, but the integrand is no longer a continuous function of time. To make the optimisation problem admissible, we choose to approximate the indicator function in the integrand by a smooth bump function, $\Phi$, which we define as 
\begin{equation}
    \Phi(t) = \frac{1}{1 + \exp\left(- \frac{t - t_1}{\epsilon}\right)}\cdot \frac{1}{1 + \exp\left(- \frac{t_2 - t}{\epsilon}\right)},
\end{equation}
where $\epsilon$ determines the length scale of the region where the function $\Phi$ transitions from zero to one. We set $\epsilon = 50^{-1}$ so that $\epsilon \ll t_1$ in all of our considered optimisation problems. Now we obtain an admissible optimal control formulation, given by
\begin{equation}
    E(v) = \int_0^T \Phi(t)\cdot(v-v^\star)^2 \dd t, \label{chapter3_eq:ConstantVelocityOnIntervalProblem}
\end{equation}
which yields a modified version of the adjoint equations for $\lambda_v$ and $\lambda_{s_{\text{eff}}}$, which are Equations~\eqref{chapter3_eq:constantVelocityAdjointV} and \eqref{chapter3_eq:constantVelocityAdjointSeff}, respectively. These become 
\begin{align}
    \frac{\dd \lambda_v}{\dd t} &=\Phi(t)\cdot2(v - v^\star) + \gamma \lambda_v, \label{chapter3_eq:ConstantVelocityOnIntervalAdjointV}\\
    \frac{\dd \lambda_{s_{\text{eff}}}}{\dd t} &=  - \alpha \lambda_v + \tau_e^{-1} \lambda_{s_{\text{eff}}}, \label{chapter3_eq:ConstantVelocityOnIntervalAdjointSeff}
\end{align}
subject to the terminal conditions $\lambda_\bullet(T) = 0$. Note that, in contrast to the optimal control problem for constant velocity in the previous subsection, there is no need to set up a terminal condition for the velocity. This means that the zero initial conditions of Equation~\eqref{chapter3_eq:SimpleInitialCondition} can be imposed. As before, we solve this boundary value problem using the bvp4c numerical scheme introduced in Section~\ref{chapter3_section:simpleOptimalControl}. 

\begin{figure}
    \centering
    \includegraphics[width=\linewidth]{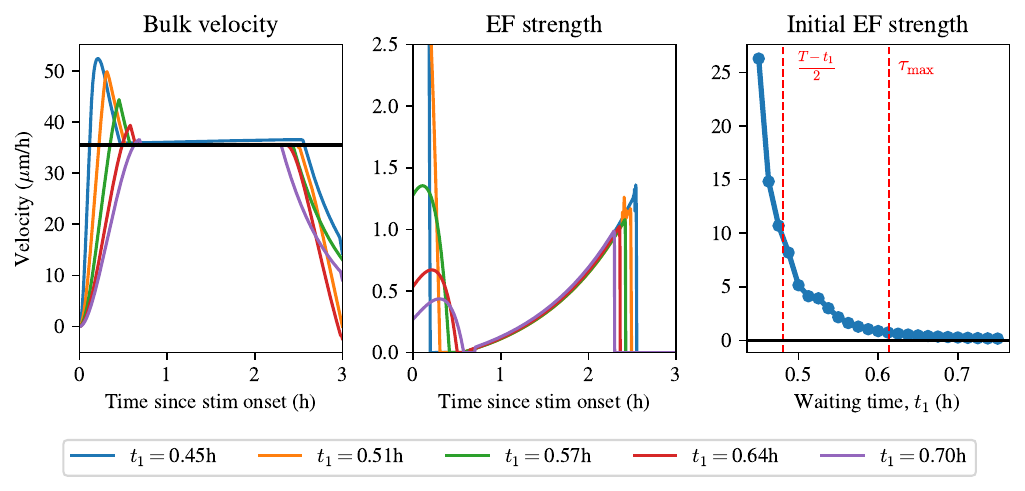}
    \caption{Optimal control to keep velocity constant in a small time window. Left: bulk velocity traces using different values of $t_1$ for stimulation. Middle: electric field strengths with different values of $t_1$. Right: initial electric field strength in the optimal control solution as a function of $t_1$.}
    \label{chapter3_fig:ConstantVelocityOnInterval}
\end{figure}

To determine the range of admissible values for the interval $[t_1, t_2]$, we make the assumption that $t_2 = T-t_1$, \textit{i.e.}  the stimulation interval is symmetric within the interval $[0,T]$. \textit{A priori}, $t_1$ should be chosen sufficiently large that a physically realistic electric field is optimal, but not so large that the field is off at $t = 0$. To understand how to choose $t_1$, then, we note that there exist two characteristic times in this optimal control problem. First, since constant stimulation for times longer than the adaptation time scale, $\tau_a$, results in a waning response from the field, choices for $t_1$ that result in the stimulation time being too long, \textit{i.e.}, $(T-t_1)/2 \gg \tau_a$, will require a large value of the electric field initially to prevent cellular adaption to the field and will hence provide unphysical solutions that are not realistic stimulation protocols in practice. At the same time, one might expect that there is a time scale that dictates when the field is initially on at $t=0$h and when it is off: if the waiting time from the beginning of stimulation, $t_1$, is shorter than this critical time scale, the electric field should be turned on immediately, since otherwise the monolayer will not reach the target velocity during the optimality time window. Intuitively, this time scale should be of the same order as the time it takes for the velocity to reach its maximum under a constant electric field. We can estimate of this time scale, $\tau_{\max}$, by maximising the expression in Equation~\eqref{chapter3_eq:asymptoticSeff}, which yields
\begin{equation}
    \tau_{\max} = \frac{\log(\tau_a) - \log(\tau_e)}{\tau_e^{-1} - \tau_a^{-1}}.
\end{equation}
For $t_1 > \tau_{\max}$ one would therefore expect that the optimal electric field density is zero at $t = 0$, while for $t_1 < \tau_{\max}$ one would expect that the density is nonzero. Therefore, one would expect that choices of $t_1$ in the region 
\begin{equation}
    t_1 \in \left[\frac{T-\tau_a}{2}, \tau_{\max}\right],
\end{equation}
will yield optimal electric field protocols that are high at the onset of stimulation when $t_1$ is at the left boundary of the interval, or electric field protocols that slowly ramp up after the onset of stimulation. We also expect values of $t_1$ smaller than the values in this range to yield physically unrealistic electric field protocols, and solutions for $t_1$ larger than the values in this range to have a near-zero initial electric field strength. In Figure~\ref{chapter3_fig:ConstantVelocityOnInterval}, we show how the optimal electric field protocol changes when different values for $t_1, t_2, v^\star$ are used.

Figure~\ref{chapter3_fig:ConstantVelocityOnInterval} shows that the optimal electric field protocols change qualitatively as $t_1$ is varied. For small values of $t_1$, close to $\tau_a$, the optimal electric field protocol exhibits a sharp overshoot of the velocity prior to reaching a plateau in the desired interval. This is achieved by the optimal electric field protocols exhibiting a sharp peak initially, bringing the velocity above the target, and then letting the velocity decrease to the target. This overshoot disappears whenever $t_1 > \tau_{\max}$. In this regime, the field allows the velocity to slowly increase to the desired target velocity. All optimal electric field protocols exhibit a ramp. This effect is clearly visible in the final plot of Figure~\ref{chapter3_fig:ConstantVelocityOnInterval}, which shows that there is a sharp decrease in the initial field strength whenever $t_1 < \tau_{\max}$. We conclude that a constant velocity can be achieved without having an overshoot when the waiting time, $t_1$, is sufficiently long. Finally, we note that by performing a grid search across possible values for the target velocity, $v^\star$, we found that these results do not change qualitatively with the value of $v^\star$, as the only change is that the maximum magnitude of the optimal electric field protocol scales linearly with $v^\star$.

\subsection{Comparison to bang-bang control}
The modest gain in distance travelled by the tissue predicted when the optimal stimulation protocol is used suggests that naive stimulation by keeping the electric field at constant field density might be an effective policy. This is not surprising given that it is the policy converged upon after years of experimental iteration, and that the key variable here is when to switch off the field. This is reminiscent of bang-bang control in linear optimisation theory. Bang-bang control occurs when the Hamiltonian, $H$, of the optimal control problem is linear in the control variable, so that it disappears from the optimality condition altogether \cite{lenhart07optimalcontrol}. In our problem the Hamiltonian is nonlinear due to the $\mu_zS^2$ term included to enforce the condition that a fixed amount of charge is injected into the system. If one now assumes instead that the field is always positive, the condition could be relaxed so that instead $S$ is bounded by some maximum strength, $0 \leq S \leq S_{\max}$. When the isoperimetric constraint is removed, the Hamiltonian becomes
\begin{equation}
    H = v^2 + \lambda_v(t) \left[ -\gamma v + \alpha s_{\text{eff}}\right] + \tau_e^{-1}\lambda_{s_{\text{eff}}}(t)\left[s - s_{\text{eff}} - I \right] + \tau_a^{-1}\lambda_I(t)\left[s-I\right],
\end{equation}
with optimality condition
\begin{equation}
    \label{chapter3_eq:BangBangDerivative}
    \frac{\partial H}{\partial S} = \tau_e^{-1} \lambda_{s_{\text{eff}}} + \tau_a^{-1}\lambda_I, 
\end{equation}
so that the optimal control is given by $s^\star = s_{\max}$, whenever the right-hand side of Equation~\eqref{chapter3_eq:BangBangDerivative} is positive by the Pontryagin optimality criterion. This, we have established in the previous numerical experiments, holds whenever $t < T$. In sum, we find that naive stimulation with a pulse is the optimal stimulation policy when the condition of total injected charge is relaxed and replaced instead with an upper bound on the electric field strength.

\section{Discussion and outlook}
\label{chapter3_section:Discussion}
In this work we have developed and calibrated a canonical model for excitation-adaptation to describe the cellular response to collective electrotaxis, and we have identified that the mechanism of excitation and adaption can faithfully describe the temporal dynamics of the bulk response to electric field stimulation. We have used this model to optimise for different experimental outcomes, including maximising the total distance travelled by the tissue bulk, maximising the terminal velocity, and keeping the migration velocity constant.

Our model shows that maximising the distance travelled by the tissue bulk is possible by changing the temporal distribution of electric field strength, but it only produces a marginal increase in performance relative to stimulation with an electric field with constant field strength. In fact, this increase is well within the standard deviation of the experimental data. We related the near-optimal performance of the constant stimulation experiment to the fact that constant stimulation is the solution to a bang-bang control problem. A similar insight arises from controlling the maximum terminal velocity of the tissue, which recovers a ramp, but has an unphysically large field strength at the end of electric field stimulation. Together,  this challenges the idea in the electrotaxis community that a bulk velocity that is not constant during stimulation -- but rather exhibits a sharp peak and a rapid decrease during stimulation --is not optimal, as this velocity profile is in close agreement with the optimal solution for the maximum distance travelled for the tissue bulk. 

Most importantly, our analysis shows that it is not possible to generate a constant velocity profile during electrotaxis when the experimental time exceeds the adaptation time scale, $\tau_a$. This is reflected in the prediction that an unphysical field strength is necessary to create a constant velocity profile for the duration of the experiment. When we relaxed the constraint to allow the tissue to slowly increase to the desired velocity and then remain constant, such optimal electric field protocols only yielded velocity profiles without an overshoot whenever the stimulation time satisfied certain constraints. Together, these findings show that there exist physical limits to how we can stimulate, both in terms of duration of the experiment, and physiological constraints on electric field strength. 

In the context of collective electrotaxis, this work can be extended and applied in several possible ways. Firstly, our framework establishes a reproducible and predictive model for collective cell migration that can be used to develop new experimental approaches. Conversely, experimental data can help validate and develop the optimal control framework proposed in this work. More generally, our modelling approach can be generalised and extended to be used in other contexts where there exists a desired collective migration outcome and one must find an optimal electric field protocol to achieve this pattern of migration. This can be achieved directly in our framework by expressing the desired movement pattern as the optimal solution to a linear control problem and using the optimal control framework put forth in this work. Secondly, in this work we have only considered uniaxial electric fields, while experimental set-ups exist to apply spatially dependent fields \cite{Cohen2014GalvanotacticMonolayers, leal23coculture, song21pten}. Our framework can be extended to optimise for desired migration outcomes that vary in both time and space. For the optimisation and control of spatially varying fields, one must first develop a spatial model of collective electrotaxis that takes into account the different migratory cues that are present in the different parts of the monolayer. Such a spatially resolved model will enable the design and implementation of physically relevant applications of electrotaxis, such as those used in wound healing or steering collective migration. Finally, this work provides a blueprint for using optimal control theory in conjunction with external inputs that can be modulated in time, to control cellular collectives. Such a tunable external input might, for example, be implemented in the context of chemotaxis, which was the external signal originally modelled by Othmer and Erban \cite{erban2005signal}. By carefully setting up an optimal control problem corresponding to the maintenance and adaptation of a chemical gradient, one could aid the experimental design of microfluidics or other ways of maintaining chemical gradients to ensure desired outcomes during chemotaxis. 

\newpage
\subsubsection*{Authors' contributions}
S.M.P. conceived the project, developed the mathematical modelling, created the code for the numerical implementation, produced all figures, and carried out the numerical experiments and analysis; R.E.B. helped design, supervised and coordinated the study. S.M.P. wrote the paper, on which R.E.B., D.J.C., and I.B.B. commented and revised. All authors gave final approval for publication.

\subsubsection*{Competing interests}
We declare we have no competing interests.

\subsubsection*{Acknowledgements}
S.M.P. is supported by an EPSRC/UKRI Doctoral Training Award. I.B.B. is supported by an NSF GRFP. D.J.C. would like to acknowledge support for this work was provided in part by the National Institute of Health Award R35 GM133574-03 and the National Science Foundation CAREER Award 2046977. R.E.B. was supported by a grant from the Simons Foundation (MP-SIP-00001828).


\newpage
\bibliography{references}
\bibliographystyle{unsrt}

\end{document}


\title{Supplementary Information for Optimal control of collective electrotaxis in epithelial monolayers}

\author{Simon Martina-Perez, Isaac B. Breinyn, Daniel J. Cohen, and Ruth E. Baker}
\date{}
\maketitle

\newpage
\clearpage
\bibliography{references}
\bibliographystyle{unsrt}
